# Multi-Dimensional, Non-Contact Metrology using Trilateration and High Resolution FMCW Ladar

**Ana Baselga Mateo[1], Zeb W. Barber[1*]**

[1]Spectrum Lab, Montana State University,
PO Box 173510 Bozeman, MT 59717
*Corresponding author: barber@spectrum.montana.edu

Here we propose, describe, and provide experimental proof-of-concept demonstrations of a multi-dimensional, non-contact length metrology system design based on high resolution (millimeter to sub-100 micron) frequency modulated continuous wave (FMCW) ladar and trilateration based on length measurements from multiple, optical fiber-connected transmitters. With an accurate FMCW ladar source, the trilateration based design provides 3D resolution inherently independent of stand-off range and allows self-calibration to provide flexible setup of a field system. A proof-of-concept experimental demonstration was performed using a highly-stabilized, 2 THz bandwidth chirped laser source, two emitters, and one scanning emitter/receiver providing 1D surface profiles (2D metrology) of diffuse targets. The measured coordinate precision of < 200 microns was determined to be limited by laser speckle issues caused by diffuse scattering of the targets.

## Introduction

Frequency Modulated Continuous Wave (FMCW) ladar due to its ability to provide much higher bandwidth (100's GHz to a few THz) and range resolution than other ladar and lidar systems is gaining in interest for imaging and non-contact metrology systems [1], [2]. The sub-millimeter range resolution provided by the large bandwidth and the high sensitivity provided by coherent detection allows these FMCW ladar systems to make micron level precision range measurements at standoffs exceeding 10 m and to multiple simultaneous returns [3]. Calibration of these sources against molecular references [4] or optical frequency combs [5], [6] can provide ppm level or ppb level accuracy, respectively. Additionally, the multi-return capability of FMCW ladar allows measurement of multiple surfaces of (or through) transparent materials or obscurants. These features make FMCW ladar a good candidate for metrology and 3D imaging problems where more traditional solutions such as structured illumination or interferometers fail or provide insufficient precision and accuracy.

Regarding multi-dimensional time-of-flight laser imaging and metrology, triangulation based systems common in industrial metrology, surveying, and architecture known as laser trackers, total stations, and 3D laser scanners are the leading solution currently on the market. These systems provide 3D coordinates by providing a single range measurement and two angular measurements (elevation and azimuth) angularly at each point. To provide high accuracy these system utilize cooperative (retro-reflecting) targets to achieve high signal-to-noise, however 3D laser scanners can achieve millimeter level repeatability off of bright diffuse targets and allow much more rapid scanning and point cloud generation across the scene. Taking a different approach, we propose and experimentally demonstrate in this paper a multi-dimensional metrology system for large-standoff, non-contact measurements of passive diffuse surfaces based on trilateration principles to leverage the high range resolution, precision, and accuracy made possible by ultrabroadband FMCW chirp sources. The goal was to provide 10 μm precision and ppm accuracy in all three dimensions of a large volume, stand-off (up to 100 m), and non-contact 3D metrology system. Unfortunately, underestimated issues with laser speckle and the coherent FMCW ladar system limited the experimentally measured precisions to the 100 μm level.

## Trilateration Metrology System Concept

Trilateration is a multidimensional positioning technique that relies solely on absolute distance measurements to determine position. Trilateration is commonly used in navigation applications, being the basis for the Global Positioning System (GPS). It is also a cost-effective method used by land surveyors to calculate undetermined positions in plane coordinate systems by measuring distances to previously surveyed points.

Simplifying to two dimensions for explanatory purposes – and for the experimental demonstrations presented here – the trilateration technique is simply described by the geometric construction of intersecting circles (see Figure 1). An unknown point $x, y$ is determined by measuring the absolute distances from two known points $(x_1, y_1)$ and $(x_2, y_2)$ given as $a$ and $b$, then solving for the intersection of the corresponding circle equations.

$$\begin{aligned} (x-x_1)^2 + (y-y_1)^2 &= b^2 \\ (x-x_2)^2 + (y-y_2)^2 &= a^2 , \end{aligned} \tag{1}$$

Here, the unknown target position $(x, y)$ lies on one of the two possible intersections. An extra measurement, $e$, from a third point at $(x_3, y_3)$ would narrow the possibility of location for the target to one of the intersections. However, often the proper intersection can be determined by other knowledge such as just knowing which side of coordinate system the unknown point is placed.

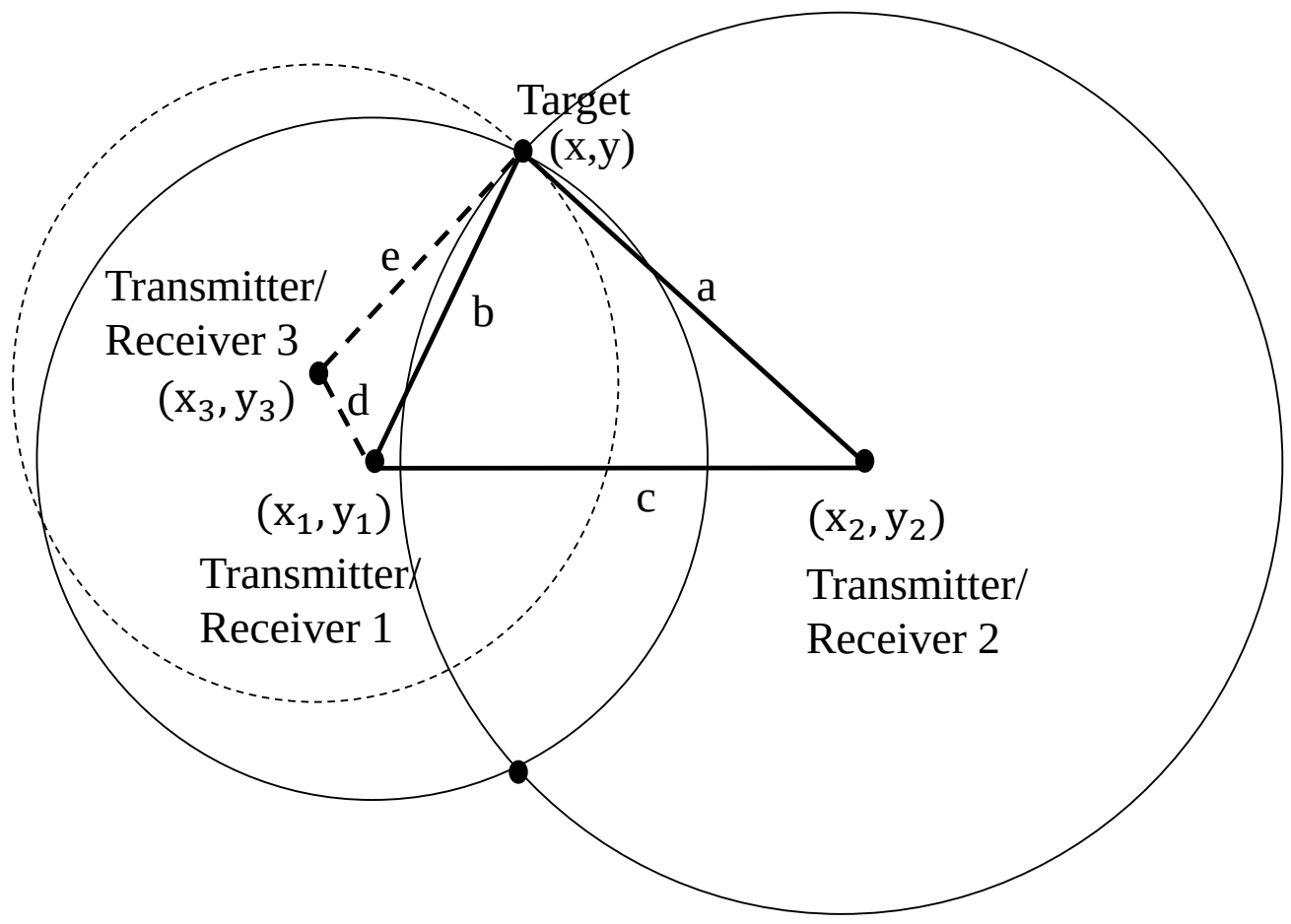

Figure 1 Trilateration technique in two dimensions.

The trilateration technique translates into three dimensions by requiring measuring distances to a minimum of three known points in 3D space and finding the intersection of spheres instead of circles. Generally, three dimensional trilateration systems (like GPS) utilize a minimum of four known points to provide redundancy and reduce volumetric uncertainty. For our system concept the unknown point is a point on a diffuse reflecting target or scene and the known points are FMCW ladar transmitters that are coupled to the same linear chirped source by use of optical fibers. The objective is then to use these sources to measure the distances $a$ and $b$. The challenges are to design a ladar based trilateration measurement system that ensures: (1) the distances being measured from each transmitter are to the same point on the diffuse target, (2) reduces the requirements on coordinated scanning of multiple transmitters/receivers (Tx/Rx), (3) provides simple calibration and/or subtraction of various range offsets caused by the scanning optics, (4) can provide self-calibration of the transmitter positions for flexible setup, and (5) has the potential to provide for eye-safe laser emission. The experimental system design described in Figure 2 attempts to overcome these challenges.

## Experimental 2D Trilateration System using FMCW Ladar

A stabilized chirped laser source [1] that is amplified with a 1 W EDFA provides an optical fiber delivered FMCW signal that is first split into a 90% path transmission to the target (Tx) and a 10% path that forms the Local Oscillator path (LO). The latter is used as a reference to mix with the returned signal (Rx) providing coherent optical gain and down conversion of the optical information to the RF regime. The Tx path is then split again with a 2x2 fiber splitter\coupler with a ratio of 99.99/0.01%. The high power path (99.99%) is then directed through a 50/50 splitter and onto the transmitters B and C producing output powers of 221 mW on each. The lowest power output (0.01%) is directed to the Tx\Rx optics labeled A with an output power of 44 µW.

For this short-range, table-top demonstration the transmitters B and C consist only of a bare single mode fiber where both emissions are manually directed to illuminate a relatively large (~30 cm diameter) common target area about 1.5 meters away. The Tx/Rx optics for A consists of an 18 mm effective focal length fiber collimator producing an ~1.7 mm Gaussian radius collimated beam. This beam is then focused to a small spot (~400 µm) on the target with an adjustable focal length positive\negative lens pair. This beam A is then reflected off and directed by a computer controlled galvo-mirror to allow scanning of the target. Care was taken to position the vertical height of the transmitters/receivers and the 1D angular galvo-scan such that the system is confined to a single horizontal 2D plane.

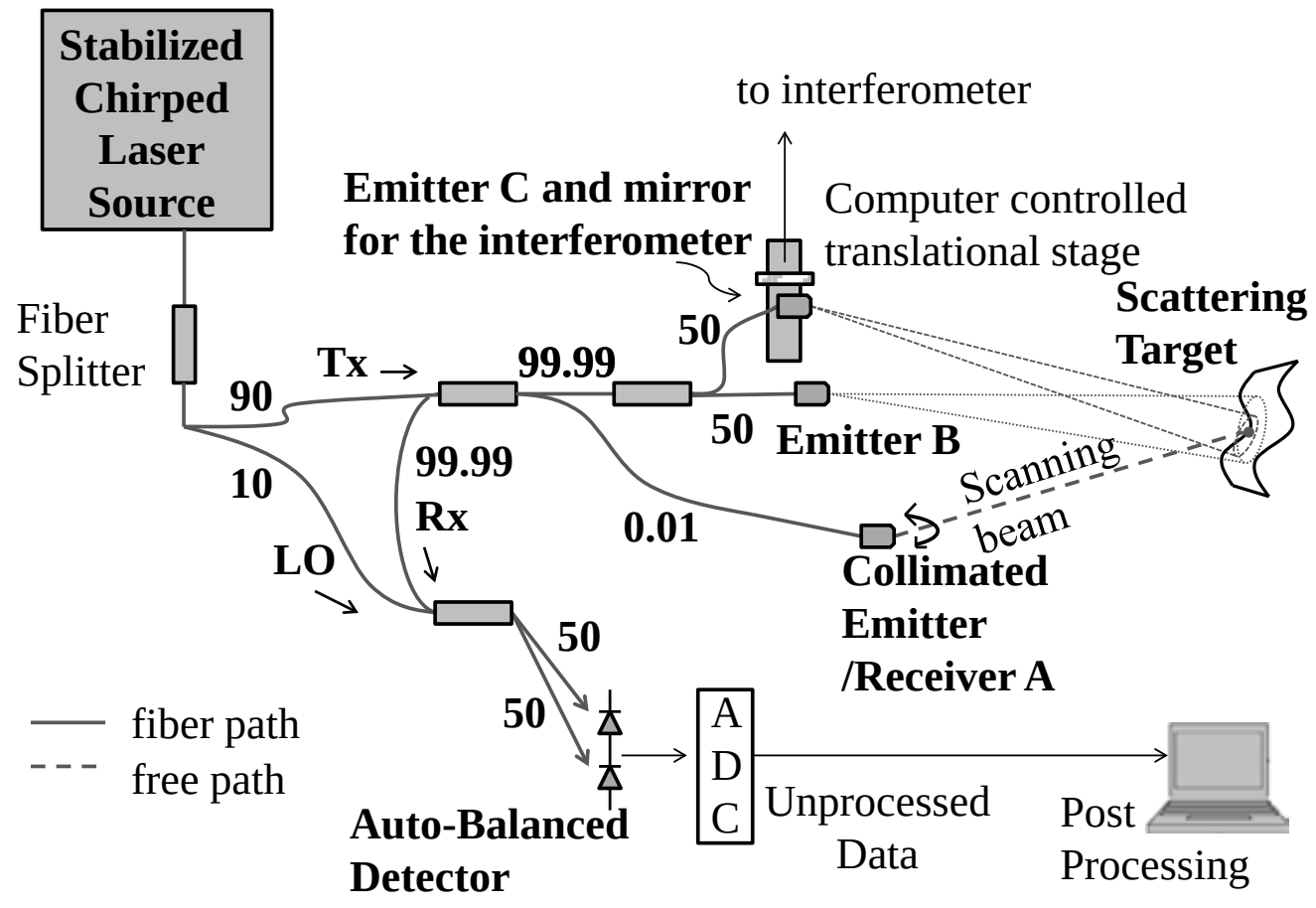

Figure 2 Experimental setup for 2D non-contact metrology of diffuse reflection targets.

The light emitted by A, B, and C that is scattered from the target and collected by receiver A is directed back through the high ratio splitter into the fourth fiber port with high efficiency. This forms the main received (Rx) signal path. Additional reflected signal from the fiber tips of emitters B and C also appears on the Rx path, but at a greatly reduced level due to the backwards coupling of the high ratio splitter. Finally, a 2x2 splitter is used to mix the Rx and LO signals onto the two ports of a homemade auto-balanced detector [7]. The output signal from the detector is captured with a NI-5122 analog-to-digital (ADC) computer card. Then, a fast Fractional Fourier Transform (FRFT) [8] is used to generate a power spectrum of the frequency region of interest (see Figure 3). The different detected signals are identified in this frequency domain, and the corresponding and desired equivalents in range calculated using the calibrated chirp rate of the FMCW signal. All signal processing was performed in Matlab.

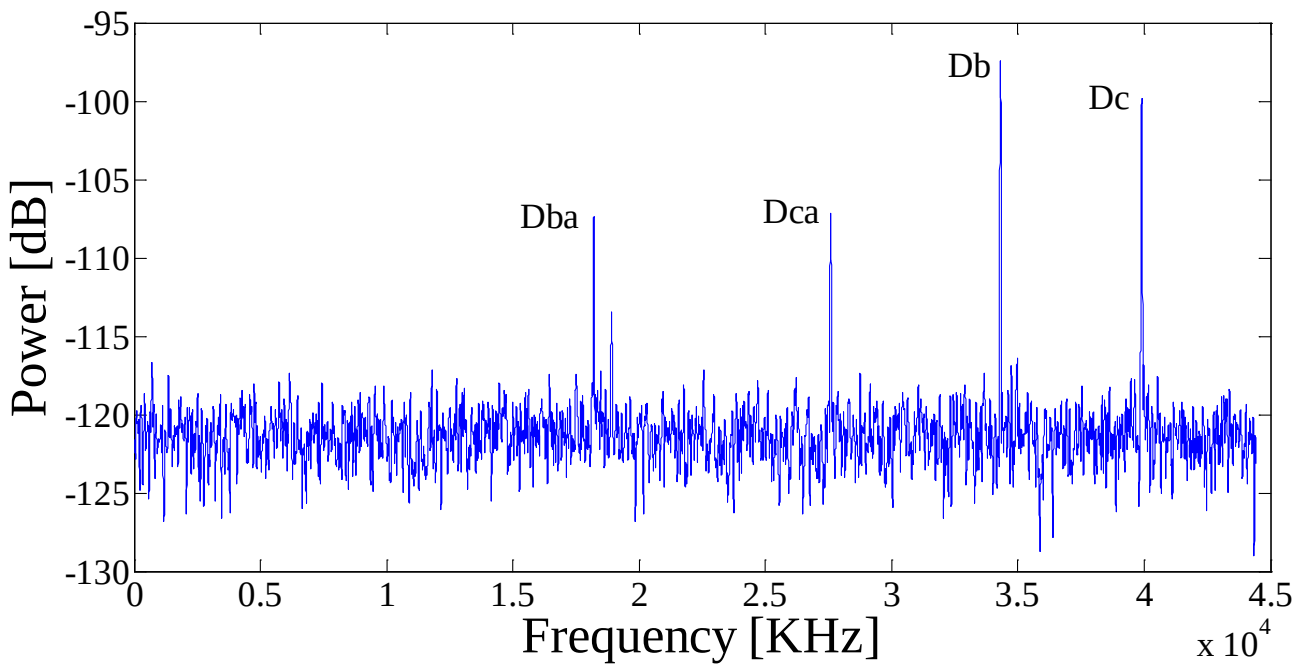

Figure 3 Detected signals in frequency domain. The corresponding and desired equivalents in range are obtained using the calibrated chirp rate (see Table 1 for nomenclature clarification).

The Rx signal measured on the single detector contains all the distance information required to both measure the free space round trip distances from Tx emitters B and C, to the target collection point, and back to the receiver A; and the fiber lengths connecting the transmitters and receivers. Figure 4 defines the different free space distances and fiber lengths in the setup, and Table 1 summarizes how the signals observed in the frequency profile Figure 3 of the bistatic ladar are used to calculate the final actual distances of interest.

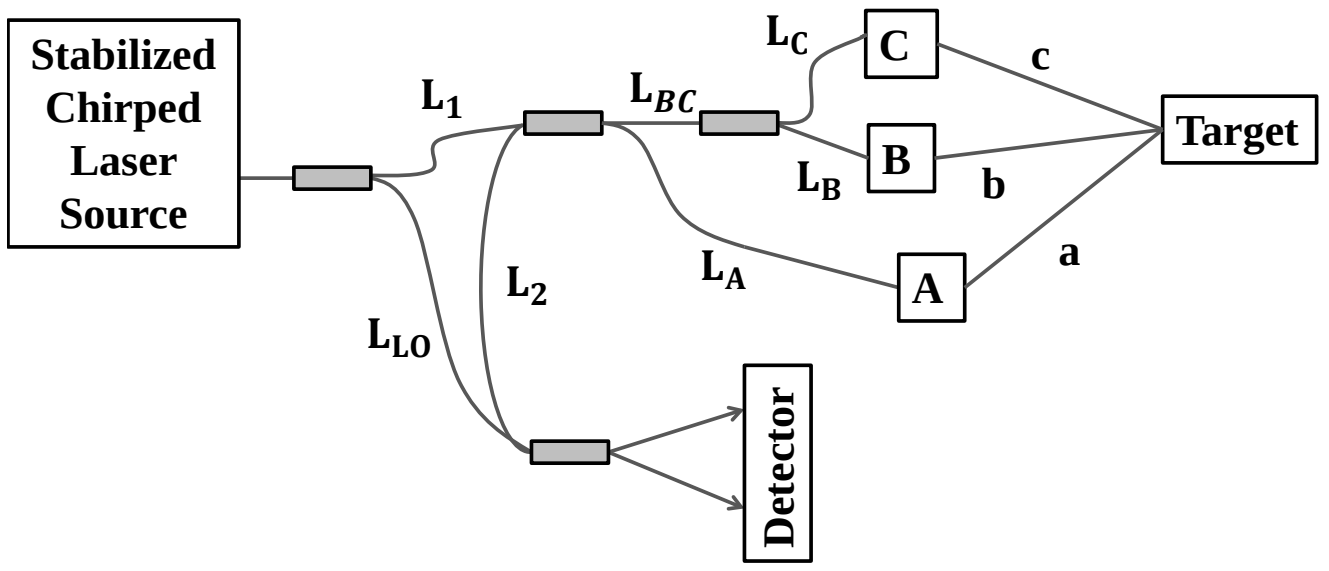

Figure 4. Diagram defining the three emitters (A,B,C) and 1 receiver (A) and defining all the relevant distances including the fiber lengths $L_n$ and the free space distances *a*,*b* and *c*.

**Table 1. Resulting range R from the trilateration setup. $L_{net}$ is defined to be $L_1 + L_2 - L_{LO}$.**

| Signal Name | Emitted by | Received by | Range measured |
|---|---|---|---|
| $D_{aa}$ | A | A | $2L_A + 2a + L_{net}$ |
| $D_a$ | Reflection from A fiber tip | | $2L_A + L_{net}$ |
| $D_{ba}$ | B | A | $L_{BC} + L_B + b + a + L_A + L_{net}$ |
| $D_b$ | Reflection from B fiber tip | | $2L_{BC} + 2L_B + L_{net}$ |
| $D_{ca}$ | C | A | $L_{BC} + L_C + c + a + L_A + L_{net}$ |
| $D_c$ | Reflection from C fiber tip | | $2L_{BC} + 2L_C + L_{net}$ |

Then, the desired distance measurements are obtained as:

$$b = D_{ba} - \frac{1}{2}D_b - \frac{1}{2}D_{aa}$$
$$c = D_{ca} - \frac{1}{2}D_c - \frac{1}{2}D_{aa} \qquad (2)$$

As shown below, the target's position is fully determined by the absolute distances $b$ and $c$, and the known positions of the emitters B $(B_x, B_y)$ and C $(C_x, C_y)$ with respect to the 2D coordinate system of choice. Therefore, the possible error on the distance $a$ due to the scanning movement of emitter/receiver A does not affect the results.

After measuring $b$ and $c$, the coordinates of the spot on the target collected by A can be calculated by Eq. 1 if the coordinates of the transmitters $B_x$, $B_y$, $C_x$ and $C_y$ are known. Using the galvo-mirror to scan the target with the focused beam from transmitter/receiver A permits taking measurements at several positions $(P_x, P_y)$ giving a profile of the surface.

Figure 2 shows that transmitter C is mounted on a computer controlled translation stage. Additionally, it indicates that the position of the stage was monitored by an optical interferometer. The purpose of this stage interferometer was to allow the position of C to be moved in a controlled manner as a method to self-calibrate the relative positions of the transmitters C and B. The self-calibration proceeded as follows: C and B commonly illuminate an area on the target and a collection scan position of A was chose and held steady, while the distance data $b_i, c_i$ was collected from a series of points defined along the motion of the stage as $C_{x_i}, C_{y_i}$. Using the set of $c_i$ distances and the known transmitter positions the steady scan point coordinate $(P_x, P_y)$ can be calculated, and from this now known point the position of transmitter B can be estimated. Several such calibrations can be performed for different scan points P to improve the estimate of the coordinates $B_x, B_y$. Unfortunately, due to the limited motion of C this self-calibration method could only provide millimeter level calibration of the relative position of B. For a true metrology system, the calibration of the relative positions of the transmitters B and C will have to be at least as good as the desired measurement accuracy. Other calibration methods were considered including using a known flat surface as a calibration target, or aligning the collection scan point of A onto two widely separated but well known calibration points, or directing the transmitters toward one another to directly measure their separations.

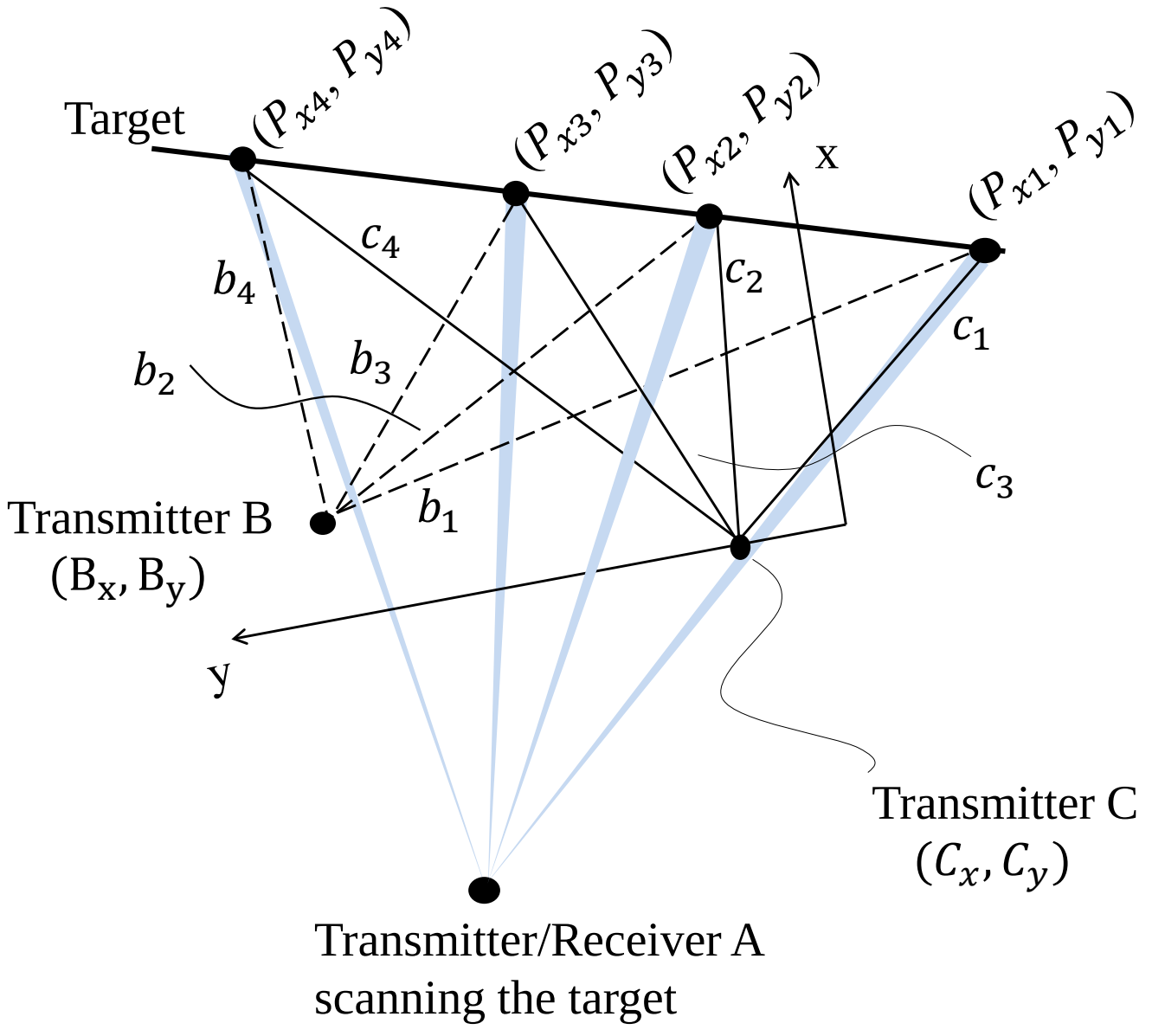

Figure 5. Diagram showing the 2D geometry of the experimental demonstration setup to horizontally scan the target and get a 2D profile of its surface.

This innovative configuration of fiber connected transmitters and receivers serves several purposes. First, by utilizing a bistatic Tx/Rx approach (unlike most ladar and lidar systems that use monostatic designs) one gains flexibility in the relative spot sizes and divergences for the Tx and Rx optics. In particular, while emitters B and C broadly illuminate the diffuse target only light collected by the separate receiver A is actually measured. As only light received by A is measured, the point on the target can be confined to a small area by utilizing receiver optics in a confocal collection geometry helping to solve challenge (1). This also relaxes the beam pointing, size, and divergence requirements on the transmitters B and C helping to solve challenge (2). In addition, the high optical power emitted by B and C can be made eye-safe due to its larger beam size and high divergence, while the small beam size and low divergence light from A is weak helping to solve challenge (5). Second, the high ratio splitter simultaneously distributes light among the high power Tx paths B,C and low power Tx/Rx path A; and acts as an efficient circulator directing nearly all of the light received by A along the Rx path to efficiently reach the heterodyne detection maximizing the SNR of the signals. It also multiplexes all of the relevant signals needed to calibrate out the fiber paths and free space path $a$ helping to eliminate common mode errors and reducing optical frontend and electronic backend resource requirements. Finally, the system design should allow self-calibration of the transmitter positions through carefully designed calibration procedures to allow flexible setup and configuration of the metrology system.

## Experimental Results

The FMCW laser source used for the results presented here was a Bridger Photonics SLM-M broadly tunable (~3 THz) chirped laser [9] that was loaned to us for these measurements by Bridger Photonics Inc. This laser is actively frequency stabilized to a fiber interferometer using the techniques described in Ref [1]. For these experiments the chirp repetition rate was 30 Hz and approximately 8 ms per repetition period was utilized providing 1.6 THz bandwidth and a Fourier window limited range resolution 94 µm. The lowish SNR due to large area flood illumination of the target by transmitters B and C resulted in a range precision of ~10 µm for when the scan position of receiver A was held steady on the target.

However, even for very small relative motions between the transmitters and the target, we noticed significant fluctuation in both the returned signal power in the range peaks and the measured range. Additionally, it was sometimes noticed that one peak or more would show up with a double peak. These effects are attributable to laser speckle, which in single mode coherent imaging systems can causes very strong amplitude and phase modulation effects with only wavelength scale changes in motion. The NIST group provided an analysis of these effects for FMCW ladar measurements in Ref. [10], showing that sub-10 µm measurements were still possible. A couple significant difference between their analysis and our measurement geometry is that they assumed a monostatic collection geometry and illumination roughly normal to the target surface. In our system, neither of these are true and our results show significantly more variance than expected by that analysis. In the case of bistatic geometries with illumination and collection far from normal, we empirically observed that the variance was roughly proportional the range extent of the target within the Gaussian beamwidth of the collection spot. This makes intuitive sense as any scatterer or collection of scatterers in the collection spot can dominate the signal measured by the coherent system, pulling the central measured range. While aware of the potential issues with laser speckle prior to observing these effects, in hindsight a simple analysis would have given an estimation of the size of these effects.

To help mitigate some of these effects (particularly the double peak issue) we found it useful to divide the collected time domain data into three overlapping portions consisting of the first half, middle half, and second half of the signal, the apply a separate raised cosine Fourier window to each piece calculating the individual FRFTs of the pieces and incoherently averaging the result. In this way, each piece of data provides a slightly different realization of the speckle providing some speckle averaging. However, this comes at the expense of reducing the range resolution again by two. The center of mass of the peak was used to calculate the center range. To greatly improve on the current precisions, a full analysis of the speckle effect in coherent ladar for bistatic non-normal incidence measurements will be required to fully understand and provide mitigation of these effects.

The first target presented here was a machined, black-anodize aluminum plate. The target was placed about 1.5 m from the transmitters at a slight angle relative to the central axis. This target was used first to perform the self-calibration procedure described above. Then a high density profile of the plate was collected using the galvo-mirror to slowly scan across the surface. The profile results are shown in Figure 5. The coordinate system was defined by assigning the origin as the home position of the C transmitter on the computer controlled stage and defining the y-axis as the direction of the motion of this stage. The x-direction was then perpendicular to this pointing in the main direction of separation between the target and the measurement system (i.e. x roughly equates to range). This data set was used to investigate the precision of the measurement system. The residual absolute distance from the measured points to the linear fit line of the target were obtained using basic geometry principles and are displayed in a histogram in Figure 6. The RMS of these transverse residuals is found to be 111 µm, which is below the range resolution for this source, but almost 10 times the result obtained when looking at a steady point on the target showing the deterioration due to speckle.

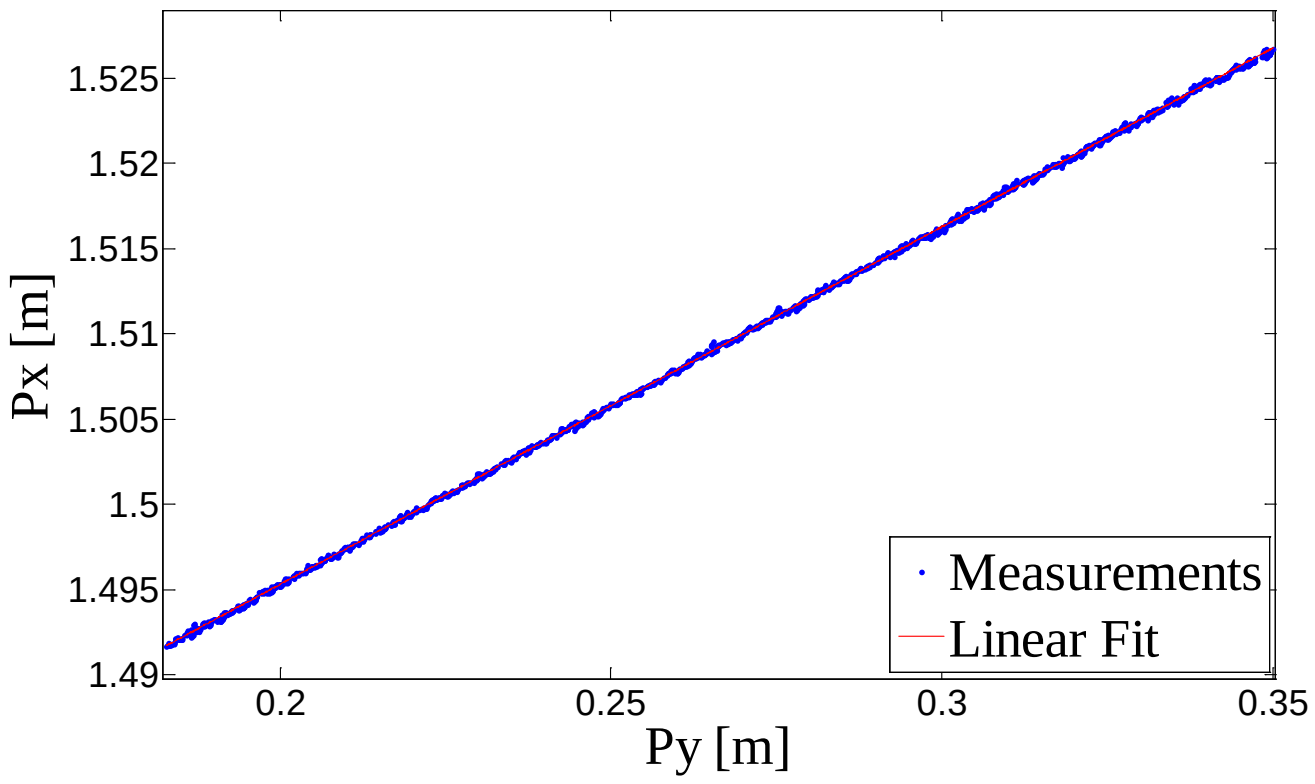

Figure 6. Scanning results for a flat non-cooperative target.

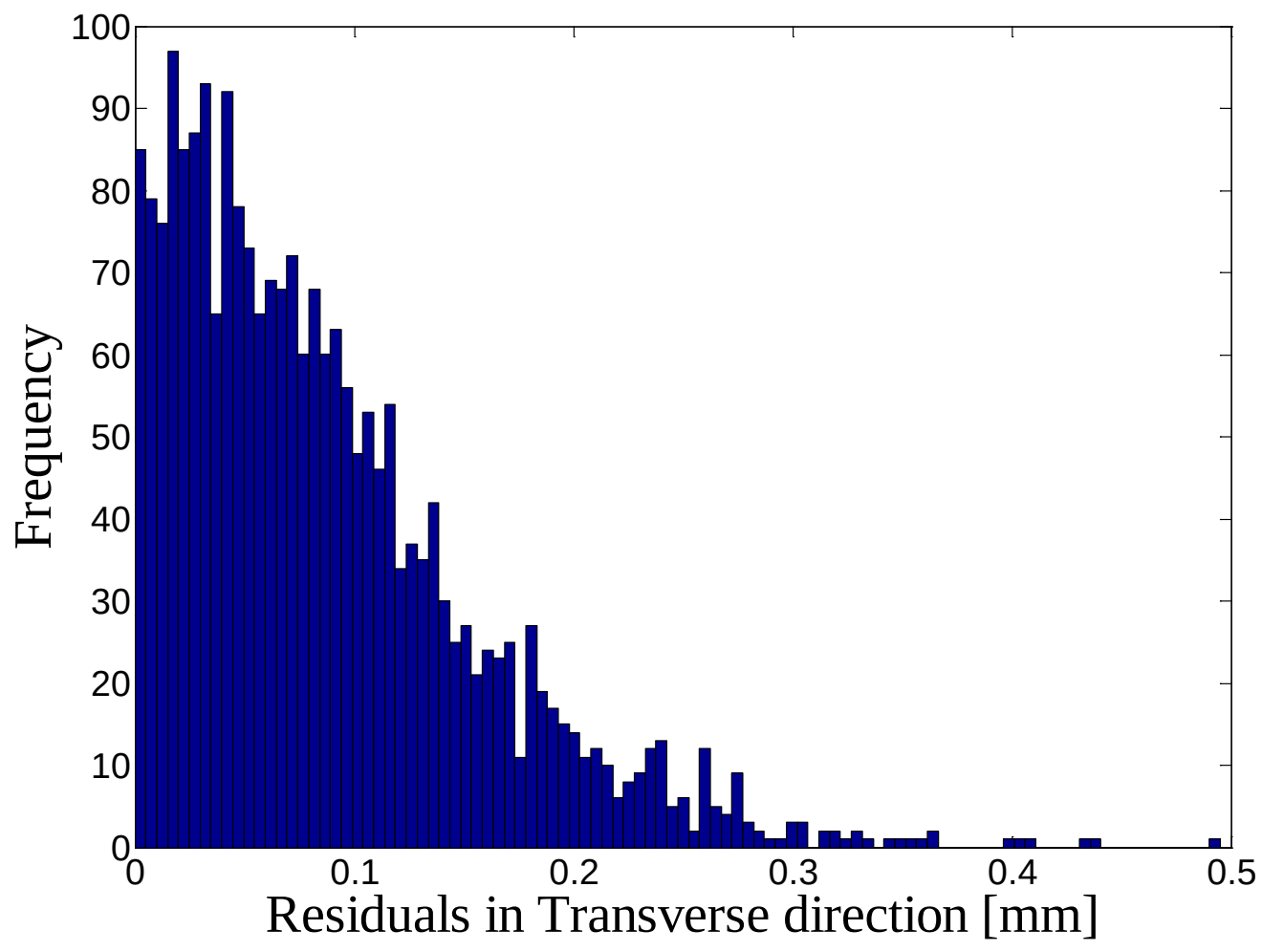

Figure 7. Histogram of the residuals in the transverse direction after a linear fit of the scanned profile.

As a second demonstration, we measured a 2D profile of a custom target machined with different features of various sizes (see Figure 8). To extract the underlying features of the 2D profile and estimate the residuals, a Savitzky-Golay method was used to smooth the data. The extracted features show good agreement with the known geometry – at least to the level of the speckle dominated residuals – and demonstrates the ability to elegantly handle various types of features such as sharp rectangular incisions. The residuals show a standard deviation of 190 μm, which is consistent with the results from the flat plate.

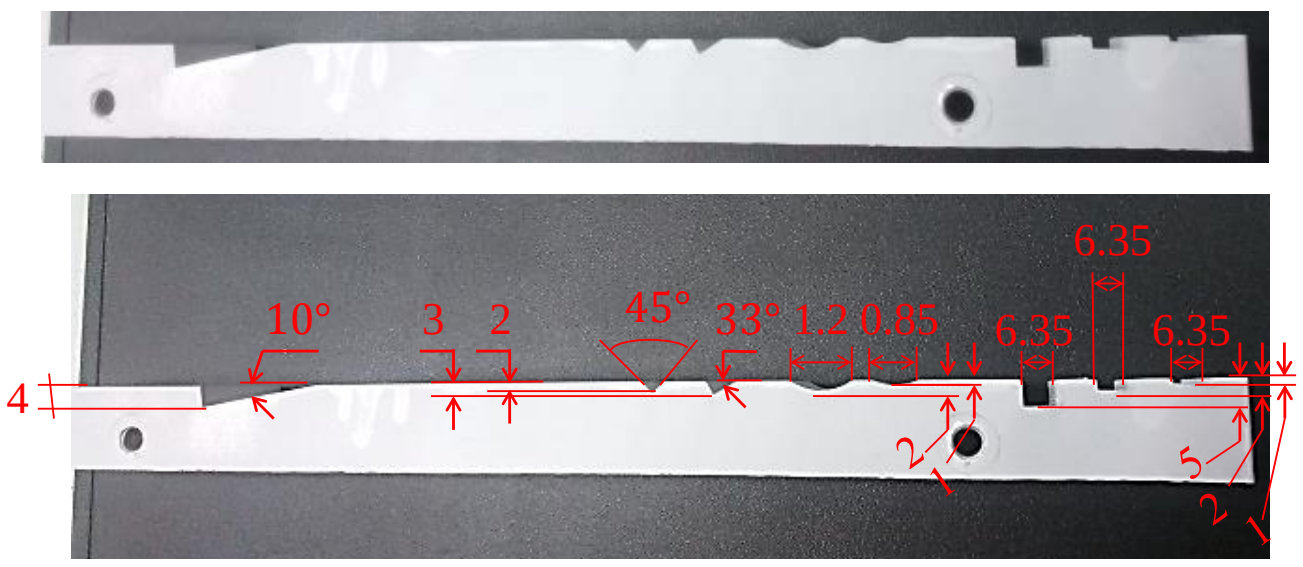

Figure 8. Machined aluminum plate with various features with its dimensions in mm (red).

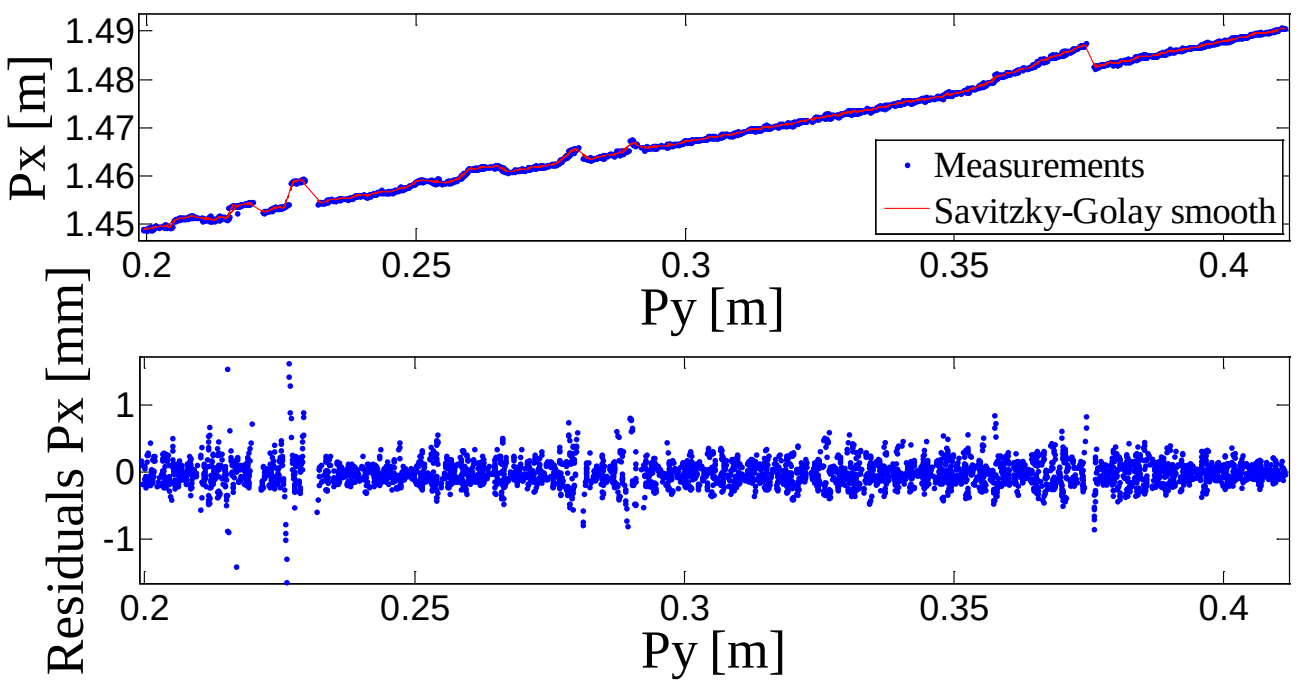

Figure 9. 2D profile results with residuals from the Savitzky-Golay smooth.

## Discussion

We have demonstrated a two dimensional metrology system for diffuse targets based on trilateration principles and using FMCW ladar to make non-contact range measurements. The 2D profile results presented above show a coordinate precision in all dimensions of ~100 to 200 μm at a stand-off of 1.5 m. However, we believe this method is easily extendable to standoff's exceeding 10 meters and 3D coordinate measurement, merely requiring the addition of a single transmitter and placing the transmitters appropriately for the longer range (i.e. providing large transmitter separations). Finally, while we demonstrated a self-calibration method using controlled movement of one of the transmitters, improved calibration will be required to ensure full accuracy to go along with the coordinate precision.

As discussed above, the measured coordinate precision is not limited by range resolution or photon budget (SNR) of the system, but rather by speckle effects which are proportional to the spot size on the target. The speckle limited precision would improve proportionally with smaller collection spots on the target, however this requires increasing the size of the collection optics to decrease the diffraction limited transverse resolution. At 10 m standoff to keep similar precision would require optics of >25 mm diameter, and increasing linearly with standoff. At some point, increasing the size of the collection optics is impractical and it might payoff to utilize synthetic aperture ladar (SAL) imaging techniques enabled by the coherent FMCW ladar technique [11] to provide improved transverse resolution of the collection spot. Additionally, speckle mitigation solutions like using several transmitters and/or receivers placed at different locations to provide speckle averaging could also be of benefit.

While these results are competitive with the precisions from commercial 3D laser scanners (~1 to 3 mm), they show that meeting the goal of 10 μm precision and accuracy in all dimensions for stand-offs up to 100 m will be very difficult due to the speckle related pulling effects. Comparing the proposed trilateration solution (with the current speckle limited precision) to a potential FMCW ladar metrology solution based on the more traditional triangulation based methods, we find limited benefit. Metrology grade commercial rotary encoders provide superb few milliarcsecond angular resolution and accuracy of 5 microradians. While this limits the transverse coordinate accuracy to 50 μm at 10 m range (~5 ppm) in a triangulation based solution, the speckle issues should be similarly limiting for both systems if the target object geometry does not provide a near normal angle of illumination. In conclusion, high resolution FMCW ladar has great potential to improve non-contact, large standoff 3D metrology. However, mitigation of the speckle issues will be required to make our proposed trilateration solution an improvement over a triangulation based approach.

## Acknowledgements

The authors would like to thank Bridger Photonics Inc. for the loan of their SLM-M laser and for discussions about laser speckle, precision measurements, and calibration issues. This work was supported by MME/GOALI program NSF Grant #1031211. A more detailed description of this work is contained in Ms. Baselga Mateo's Master's Thesis [12].